\documentclass[sn-mathphys-num]{sn-jnl}

\usepackage{graphicx}%
\usepackage{amsmath,amssymb,amsfonts}%
\usepackage{amsthm}%
\usepackage{mathrsfs}%
\usepackage[title]{appendix}%
\usepackage{xcolor}%

\usepackage{textcomp}%
\usepackage{booktabs}%

\usepackage{algorithm}%
\usepackage{algorithmicx}%
\usepackage{algpseudocode}%
\usepackage{listings}%

\def\name{TrackEff}

\DeclareMathOperator{\atan2}{atan2}

\begin{document}

\title{Estimating the track-reconstruction efficiency in phenomenological proposals of long-lived-particle searches}

\author{Emilie Bertholet}
\email{ebertholet@tauex.tau.ac.il}

\author{Abner Soffer}
\email{asoffer@tau.ac.il}

\affil{School of Physics and Astronomy, Tel Aviv University, Israel}

%\author[a]{Emilie Bertholet,}
%\affiliation[a]{School of Physics and Astronomy, Tel Aviv University, Tel Aviv 69978, Israel}

%\author[a]{Abner Soffer}
%\emailAdd{asoffer@tau.ac.il}

\abstract{
Phenomenological proposals of searches for new, long-lived particles face a challenge when estimating the reconstruction efficiency of displaced charged-particle tracks.
The efficiency depends not only on the charged-particle's momentum vector, but also on its production point in relation to the detector boundary and substructure elements. 
Phenomenological studies generally do not have access to GEANT4-based simulation, which is used for calculating the efficiency in experimental analyses. 
To address this need, we have developed \name, a python software package for estimating the track-reconstruction efficiency. 
The default detector geometry within \name\ is that of the Belle~II drift chamber. 
However, tunable parameters enable modification of the geometry to correspond to other tracker configurations. 
\name\ uses a simplified model of the tracker to determine the number of detector hits associated with each track.
The user can then decide whether the track would be detected based on the number of hits, potentially in association with any other information.
}

\maketitle

%%%%%%%%%%%%%
\section{Introduction and motivation}
\label{sec:intro}

Recent years have seen a growing interest in searches for new, long-lived particles (LLPs) that arise from physics beyond the standard model, at a variety of experiments.
Here we focus on LLPs that decay away from the beam collision point yet inside the tracking volume of the detector.
In this case, the tracks of the charged daughters created in the LLP decay emanate from a displaced vertex (DV).
The presence of a DV in the signal signature usually provides highly effective background suppression, making such searches experimentally attractive~\cite{Lee:2018pag}.

When phenomenologists propose a DV-based search, they are faced with the difficult task of estimating the efficiency of the DV reconstruction as a function of its position and other parameters.
While the average efficiency for reconstructing promptly produced tracks can be roughly obtained from many published experimental papers, this is generally not the case for a displaced track that originates from a DV.
As a result, some phenomenology papers ignore question of efficiency altogether, while others use a simplified parameterization to estimate it.
Either approach results in an imprecise estimation of the predicted sensitivity of the proposed search.
This calls for a tool for estimating the tracking efficiency that is based on generator-level Monte Carlo without the use of full detector simulation, while accounting for the basic geometric features of the detector.

Another use case of such a tool is the early stages of a new detector concept. 
Construction of a full GEANT4~\cite{GEANT4:2002zbu} model is time consuming. 
By contrast, a tool that enables easy construction of a simplified detector model can facilitate quick, albeit less accurate, tests of the performance that results from the detector geometry. 

To address these needs, we have developed the python-based package \name, which is available for use from Zenodo~\cite{bertholet_soffer_2024}.

\name\ can be used for a variety of detector configurations.
Nonetheless, since it was developed with the Belle~II experiment in mind, we focus on Belle~II in what follows. 
Belle~II~\cite{Belle-II:2010dht,Belle-II:2018jsg} is an asymmetric-energy $e^+e^-$ $B$-factory experiment currently operating at SuperKEKB, the worlds highest luminosity collider. 
With electron and positron beam energies of 7 and 4~GeV, respectively, the experiment typically runs at a center-of-mass energy $\sqrt{s} = 10.58$~GeV, corresponding to the $\Upsilon(4S)$ resonance.
Typical studies at Belle~II utilize the production of $B\bar B$, $\tau^+\tau^-$, or $c\bar c$ pairs, as well as other bottomonium or hadronic states.
Several DV-based LLP searches have already been conducted at Belle II~\cite{Belle-II:2023ueh} and at its predecessor $B$~factories, the BABAR~\cite{BaBar:2015jvu} and Belle~\cite{Belle:2013ytx, Belle:2024wyk} experiments.
Directly related to \name\ are proposals for LLP searches at such experiments~\cite{Dey:2020juy,Bertholet:2021hjl,Dib:2019tuj,Ferber:2022ewf,Zhou:2021ylt,Kang:2021oes,Duerr:2020muu,Duerr:2019dmv,Filimonova:2019tuy,Jaeckel:2023huy,Bandyopadhyay:2022klg,Ferber:2022rsf}.

In what follows we describe the algorithm used in \name\ in Section \ref{sec:eff}.
Section \ref{sec:validation} describes the method by which the algorithm was validated and its parameters were tuned.

%%%%%%%%%%%%%
\section{Estimation of the track reconstruction Efficiency}
\label{sec:eff}

\subsection{Basic idea}
\label{subsec:basic}

The reconstruction efficiency is estimated based on the principle that a charged particle will be reconstructed if it left a sufficient number of hits in the tracker.
In the case of Belle~II, we consider the number of hits in the central drift chamber (CDC)~\cite{Belle-II:2010dht}. 
Belle and Belle~II physics analyses sometimes include an explicit track-quality requirement of 20 CDC hits (e.g., Refs.~\cite{Belle-II:2023ksq, Belle:2024wyk}).
Even without this explicit requirement, the number of hits is used in various stages of the tracking algorithm~\cite{BelleIITrackingGroup:2020hpx}.
Therefore, to within the approximations of a phenomenological paper, the presence of 20 hits is a reasonable criterion to determine whether a charged-particle track will be reconstructed.
\name\ estimates the number of hits geometrically, as described below.
It is left to the user to cut on the desired number of hits.

We note that additional considerations may reduce the efficiency to below the \name\ estimate.
In particular, the track-finding algorithm used by the experiment may not be ideal or may have difficulties dealing with large track multiplicities, hard scatters, etc.
Additional limitations may exist in vertexing algorithms used to reconstruct DVs from displaced tracks.
Given the recent development of an improved tracking algorithm for Belle~II~\cite{Reuter:2024kja}, these are temporary limitations that can be removed.
Since \name\ is to be used for proposals of new searches and detector designs, it is appropriate to consider its efficiency estimations as what can be achieved with future tracking and vertexing algorithms, even if the algorithms currently in use fall somewhat short of this performance. 

\subsection{Estimation of the number of hits}

\begin{figure}[htp]
    \centering
    \includegraphics[width=1\textwidth]{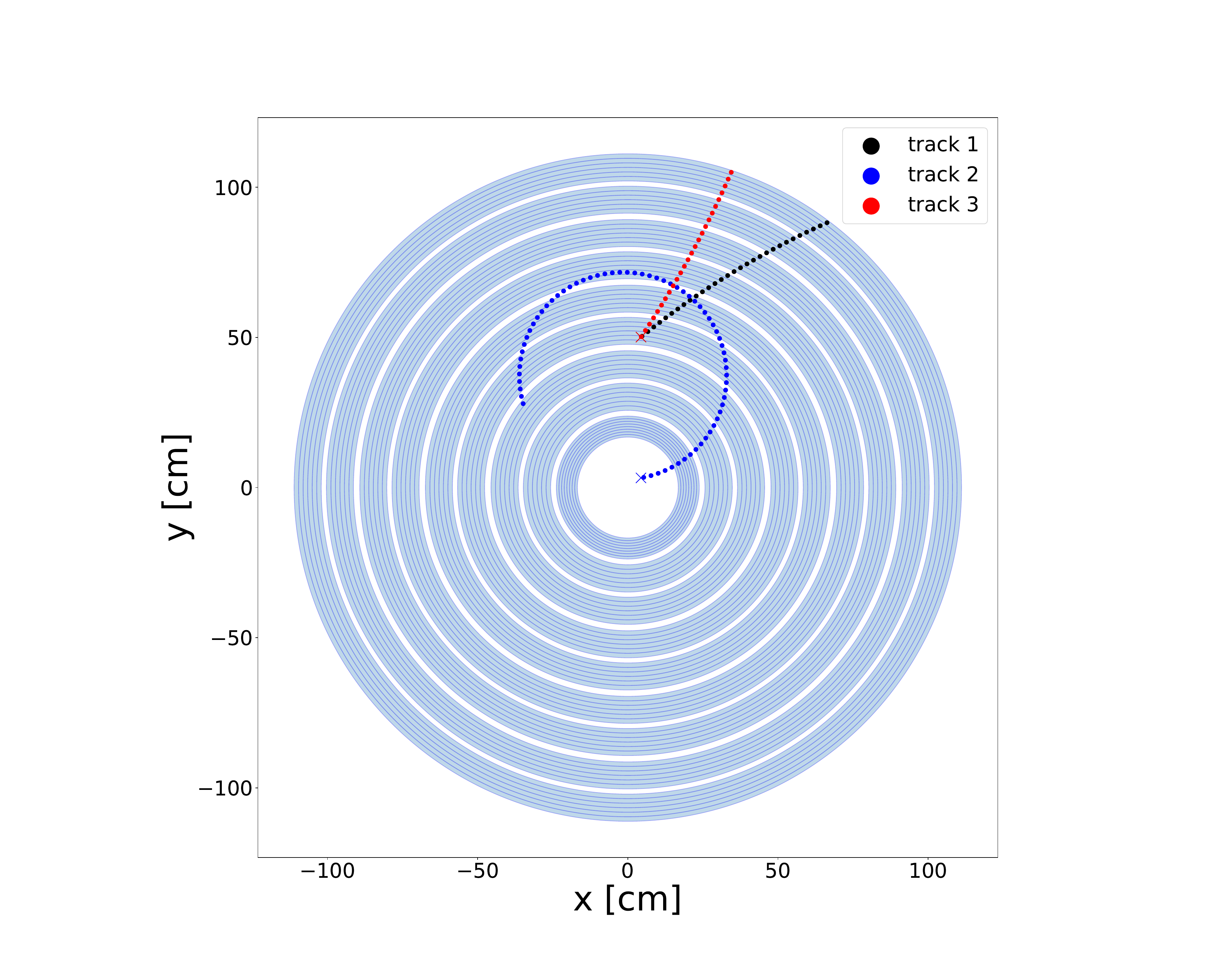}
    \includegraphics[width=1\textwidth]{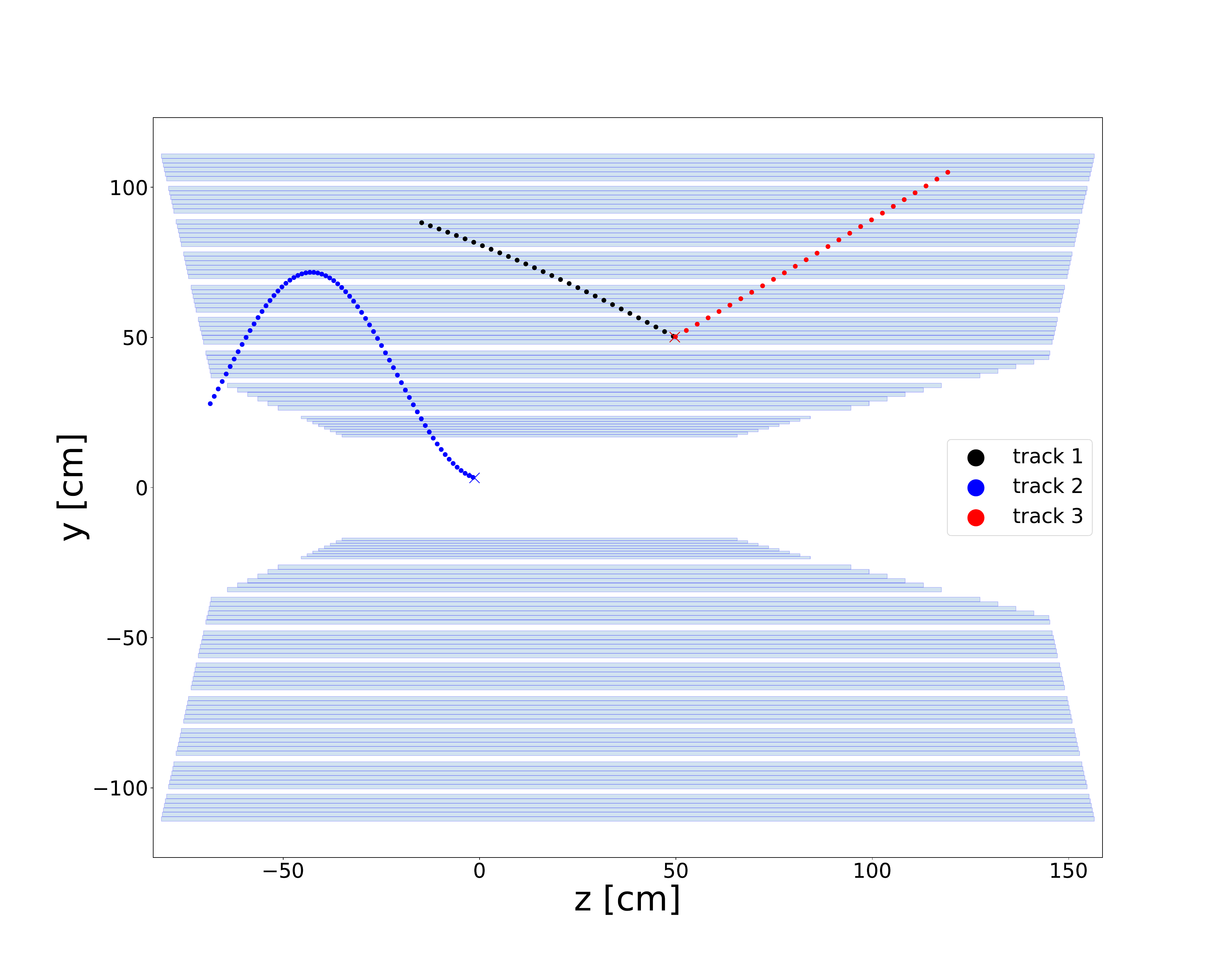}
\caption{Illustration of the CDC model overlaid with three tracks in $xy$ and $zy$ views. 
Each layer is shown as a light-blue band, and superlayers are separated by gaps. 
Tracks are indicated as dots at  fixed arc-length steps. 
Tracks 1 and 3 originate from the vertex $V_1=(4.39, 50.1, 49.7)$~cm (marked with the red $\times$), and track 2 originates from $V_2=(4.39, 3.21, -1.27)$~cm (blue $\times$).
The charges of the tracks are $q_1=1$, $q_2=q_3=-1$, and their momenta at their production vertices are $p_1=(1.10, \, 0.89, \, -1.25)$~GeV, 
$p_2=(0.17, \, 0.03, \, -0.07)$ GeV,
$p_3=(1.00, \, 1.5, \, 2.0)$~GeV.
Tracks end at the boundary of the CDC, although this may not be apparent in the 2-dimensional views.}
\label{fig:eventDisplay}
\end{figure}

The default detector-geometry parameters of \name\ correspond to a simplified model of the Belle~II CDC based on Table~6.2 and Fig.~6.2 of Ref.~\cite{Belle-II:2010dht}. The model divides the CDC into 9 radial sections, called superlayers. 
Each superlayer is divided into a number of layers.
The innermost superlayer has 8 layers, and the other remaining superlayers contain 6 layers each. 
All the layers of a given superlayer are taken to have the same radial width.
Each layer may have a different length in the $z$ direction (along the beams).
In the default configuration, the lengths are determined from Fig.~6.2 of Ref.~\cite{Belle-II:2010dht}. 
The actual Belle~II CDC has axial and stereo layers, which enables measurement of the $z$ coordinates of hits. 
However, since this has no bearing on counting the number of hits, in the model all layers are considered axial for simplicity.
Each layer contain a specific number of azimuthal cells, varying from 160 for the layers of the innermost superlayer to 384 for outermost superlayer. 
All the cells in a given superlayer have the same azimutual width.
Thus, each cell occupies a rectangle ($\Delta r, \Delta \phi$) in $r\phi$ space.
Each cell corresponds to an anode sense wire surrounded by field wires in the real CDC.
The user can modify parameters of the CDC model, such as the number of layers, their lengths, radii, and the number of cells in each layer.
The default detector model in \name\ is shown in Fig.~\ref{fig:eventDisplay} with three example tracks overlaid.

It is envisioned that \name\ will be used with the output of a particle-level generator, such as PYTHIA~\cite{Sjostrand:2006za}, EVTGEN~\cite{Lange:2001uf}, etc.
Given a generator-level charged-particle's point of origin and momentum vector, \name\ first computes the particle's track parameters at the track's point of closest approach (POCA) to the $z$ axis given the magnetic field $B_z$, which by default is $1.5$~T. 
From there, it is possible to calculate the coordinates of any point on the particle's trajectory, parameterized in terms of the arc-length $s$ in the $xy$ plane, where $s=0$ at the POCA. 
Detailed information about the calculation is given in appendix~\ref{appendixA}.

A charged particle is considered to have produced a hit in a cell if it traverses a sufficiently long path within the cell.
To determine whether this condition is satisfied, \name\ take arc-length steps along the particle's trajectory and counts the number of steps inside each cell. 
The default step length is $\Delta s=1$~mm.
If the track traverses the cell close to its edge, the ionization signal created in the cell is weak due to the short track length inside the cell, and the ionized electrons may be too weak or take too long to arrive at the anode wire to be associated to the track.  
To take  this effect into account, a certain number of steps per cell is required in order for the cell to be counted as a hit. 
By default, the number of steps required inside the cell for the cell to register a hit is $n_s=10$ for most cells. 
However, for cells in the innermost superlayer, which have a smaller radial extent in the default detector model, the requirement is $n_s^{\rm inner}=2$ hits.
These values can be modified by the user to account for different cell sizes and even thin solid-state detectors.

%%%%%%%%%%%%%%%%%%%%
\section{Validation and parameter tuning}
\label{sec:validation}

To validate \name\ and tune its default parameters, we compared its efficiency estimate to measurements performed on full Belle~II Monte Carlo and on detector-collected data, published in Ref.~\cite{BelleIIeff}. 
That study involved identifying $e^+e^-\to\tau^+\tau^-$ events, where one $\tau$ decays leptonically and the other decays to a state with three charged pions.
The tracking efficiency is the fraction of events in which all three pions were found. 
More details about the measurement method can be found in Section~4 of 
Ref.~\cite{Allmendinger:2012ch}. 

\begin{figure}[htbp]
    \centering    \includegraphics[width=1.0\linewidth]{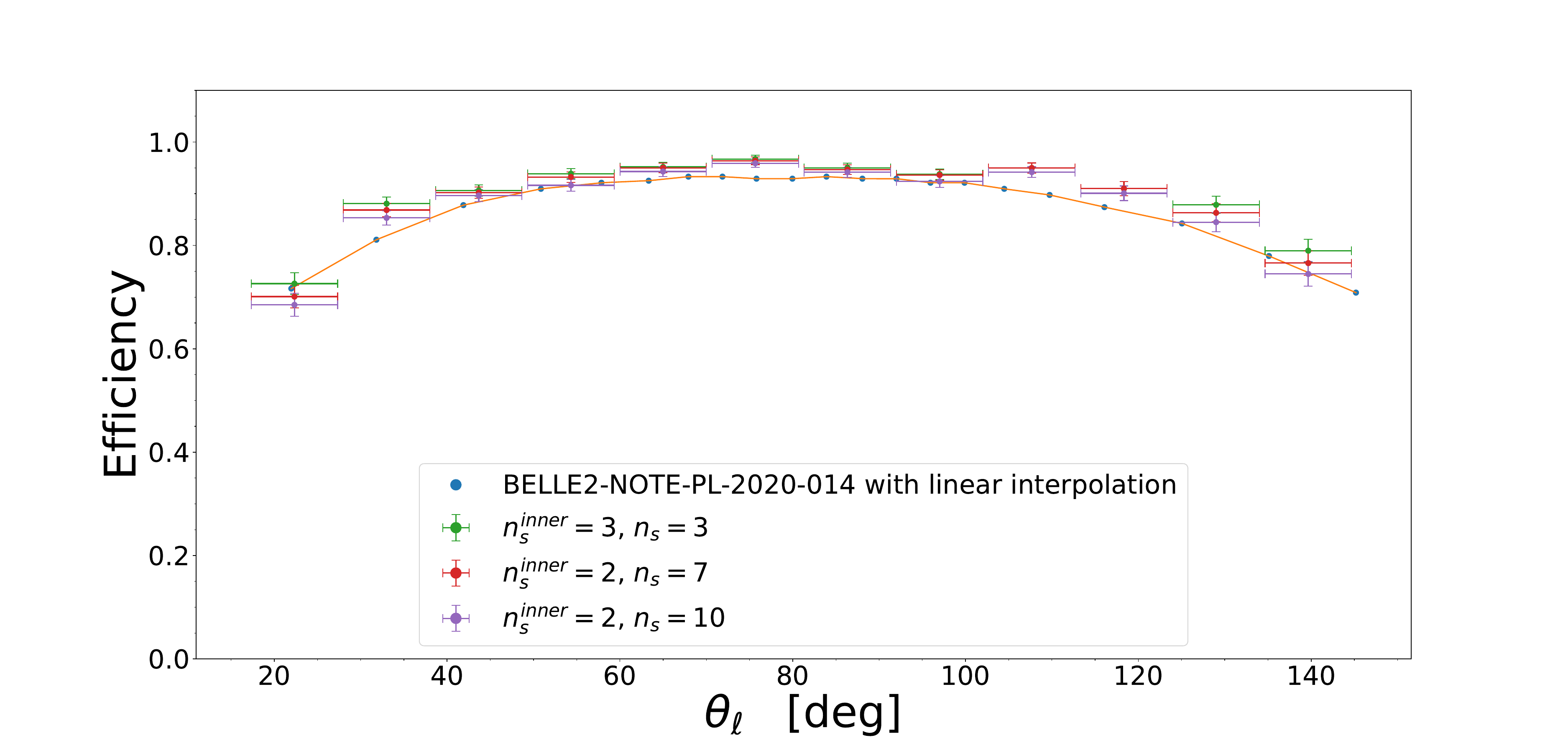}
    \caption{Comparison of the Belle~II tracking efficiency reported in Ref.~\cite{BelleIIeff} (blue circles connected by orange lines) and the efficiency estimated by \name\ (points with error bars) for different values of the number of steps, $n_s^{\rm inner}$ and $n_s$, required for the traversed CDC cell to count as a hit, when requiring 20 hits to consider a track as detected.
    Results are calculated for the pions in $e^+e^-\to \tau^+\tau^-$ with one $\tau$ decaying leptonically and the other decaying into a state containing three charged pions.
    The $x$ axis gives to the laboratory-frame polar angle of the electron or muon produced in the $\tau$ decay.}
    \label{fig:eff}
\end{figure}

In Fig.~\ref{fig:eff} we reproduce Fig.~7 of Ref.~\cite{BelleIIeff}, showing the Belle~II tracking efficiency as a function of the polar angle of the lepton produced in the $\tau$ decay (used in Ref.~\cite{BelleIIeff} since it can be measured in the data). 
This is overlaid with the \name\ estimate of the efficiency for different values of the required numbers of steps, $n_s^{\rm inner}$ and $n_s$, with all other parameters set to their default values and with 20 hits required for track detection.
As can be seen, the dependence of the efficiency on the values of $n_s^{\rm inner}$ and $n_s$ is not strong.
Nonetheless, the best match is obtained with the values of $n_s^{\rm inner}=2$, $n_s=10$, which are, therefore, taken to be the default in \name.

\section{Summary}

\name~\cite{bertholet_soffer_2024} is a publicly available python-based package for estimating the efficiency for finding a charged-particle track at a tracking detector.
The default detector-geometry parameters correspond to the Belle~II drift chamber.
However, since these parameters can be changed, \name\ can also be used for other detectors, as long as their geometries can be described within \name.  
The detector model is constructed of simplified, nearly rectangular cells in a drift chamber, but can equally be considered to be other geometries, such as drift tubes or solid-state detectors.
The package is intended for use with the output of Monte-Carlo generators.
Given the production point (or any other point along the track) of a charged particle, the package counts the number of detector layers traversed by the track and decides whether the track will be observed based on number of hits thus created.
The envisioned use cases are phenomenological papers proposing new searches for long-lived particles with a displaced-vertex signature, and studies of new detector geometries for which the time-consuming task of constructing a full-simulation model has not yet been performed.

\clearpage
\begin{appendices}
\section{Track parameters and their use}
\label{appendixA}

This appendix provides a comprehensive summary of the primary equations implemented in the software to compute the trajectory of a given particle. They are taken from Refs.~\cite{AlcarazMaestre:2021oeh} and~\cite{Belle-IIanalysissoftwareGroup:2019dlq}.

Let $\{V_x, V_y, V_z\}$ be the point of origin of a particle of charge $q$, and let $\{p_x, p_y, p_z\}$ be its momentum vector at that point. 
We calculate its trajectory assuming a uniform $B_z=1.5$~T magnetic field along the $z$-direction. 
The particle has a helicoidal trajectory that is described by five track parameters.

Using the following convenience functions
\begin{align}
\phi &= \atan2(p_y, p_x), \nonumber \\
\Delta_{\parallel} & = -V_x  \cos(\phi) - V_y \sin(\phi), \nonumber \\
\Delta_{\perp} & = -V_y  \cos(\phi) + V_x \sin(\phi), \nonumber \\
A & = 2\Delta_{\perp} + \omega(\Delta_{\perp}^2 + \Delta_{\parallel}^2), \nonumber \\
d\phi & = \atan2(\omega \Delta_{\parallel}, 1 + \omega \Delta_{\perp}), \nonumber \\
s_0 & = \frac{d\phi}{\omega}, 
\end{align}
The values of the track parameters at the point of its closest approach to the $z$ axis is then given by
\begin{align}
\omega &= \frac{q}{c}\frac{B_z}{\sqrt{p_x^2 + p_y^2}}, \nonumber \\
\tan \lambda & = \frac{p_z}{\sqrt{p_x^2 + p_y^2}}, \nonumber \\
d_0  &= \frac{A}{1 + \sqrt{1 + A \omega}}, \nonumber \\
\phi_0 &= \phi - d\phi , \nonumber \\
z_0 &= V_z + s_0 \tan \lambda, \nonumber \\
x_0 &= -d_0 \sin(\phi_0), \nonumber \\  %% check sign
y_0 &= d_0 \cos(\phi_0). %% check sign
\end{align}
From here, it is possible to get the coordinates of any point on the trajectory given the angle $\phi$ of the momentum in the $x-y$ plane or, equivalently, the $x,y$-plane arc length $s$.

\begin{align}
s & = \frac{\phi_0 - \phi}{\omega}, \nonumber \\
x & = x_0 - s \,\mathrm{sinc}(\frac{\phi - \phi_0}{2})\cos(\frac{\phi + \phi_0}{2}) ,\nonumber \\  %% check sign
y & =  y_0 - s \,\mathrm{sinc}(\frac{\phi - \phi_0}{2})\sin(\frac{\phi + \phi_0}{2}),\nonumber \\  %% check sign
z & = z_0 + s \tan \lambda
\end{align}

\end{appendices}

\bibliography{main}


\begin{thebibliography}{30}
\ifx \bisbn   \undefined \def \bisbn  #1{ISBN #1}\fi
\ifx \binits  \undefined \def \binits#1{#1}\fi
\ifx \bauthor  \undefined \def \bauthor#1{#1}\fi
\ifx \batitle  \undefined \def \batitle#1{#1}\fi
\ifx \bjtitle  \undefined \def \bjtitle#1{#1}\fi
\ifx \bvolume  \undefined \def \bvolume#1{\textbf{#1}}\fi
\ifx \byear  \undefined \def \byear#1{#1}\fi
\ifx \bissue  \undefined \def \bissue#1{#1}\fi
\ifx \bfpage  \undefined \def \bfpage#1{#1}\fi
\ifx \blpage  \undefined \def \blpage #1{#1}\fi
\ifx \burl  \undefined \def \burl#1{\textsf{#1}}\fi
\ifx \doiurl  \undefined \def \doiurl#1{\url{https://doi.org/#1}}\fi
\ifx \betal  \undefined \def \betal{\textit{et al.}}\fi
\ifx \binstitute  \undefined \def \binstitute#1{#1}\fi
\ifx \binstitutionaled  \undefined \def \binstitutionaled#1{#1}\fi
\ifx \bctitle  \undefined \def \bctitle#1{#1}\fi
\ifx \beditor  \undefined \def \beditor#1{#1}\fi
\ifx \bpublisher  \undefined \def \bpublisher#1{#1}\fi
\ifx \bbtitle  \undefined \def \bbtitle#1{#1}\fi
\ifx \bedition  \undefined \def \bedition#1{#1}\fi
\ifx \bseriesno  \undefined \def \bseriesno#1{#1}\fi
\ifx \blocation  \undefined \def \blocation#1{#1}\fi
\ifx \bsertitle  \undefined \def \bsertitle#1{#1}\fi
\ifx \bsnm \undefined \def \bsnm#1{#1}\fi
\ifx \bsuffix \undefined \def \bsuffix#1{#1}\fi
\ifx \bparticle \undefined \def \bparticle#1{#1}\fi
\ifx \barticle \undefined \def \barticle#1{#1}\fi
\bibcommenthead
\ifx \bconfdate \undefined \def \bconfdate #1{#1}\fi
\ifx \botherref \undefined \def \botherref #1{#1}\fi
\ifx \url \undefined \def \url#1{\textsf{#1}}\fi
\ifx \bchapter \undefined \def \bchapter#1{#1}\fi
\ifx \bbook \undefined \def \bbook#1{#1}\fi
\ifx \bcomment \undefined \def \bcomment#1{#1}\fi
\ifx \oauthor \undefined \def \oauthor#1{#1}\fi
\ifx \citeauthoryear \undefined \def \citeauthoryear#1{#1}\fi
\ifx \endbibitem  \undefined \def \endbibitem {}\fi
\ifx \bconflocation  \undefined \def \bconflocation#1{#1}\fi
\ifx \arxivurl  \undefined \def \arxivurl#1{\textsf{#1}}\fi
\csname PreBibitemsHook\endcsname

%%% 1
\bibitem[\protect\citeauthoryear{Lee et~al.}{2019}]{Lee:2018pag}
\begin{barticle}
\bauthor{\bsnm{Lee}, \binits{L.}},
\bauthor{\bsnm{Ohm}, \binits{C.}},
\bauthor{\bsnm{Soffer}, \binits{A.}},
\bauthor{\bsnm{Yu}, \binits{T.-T.}}:
\batitle{{Collider Searches for Long-Lived Particles Beyond the Standard
  Model}}.
\bjtitle{Prog. Part. Nucl. Phys.}
\bvolume{106},
\bfpage{210}--\blpage{255}
(\byear{2019})
\doiurl{10.1016/j.ppnp.2019.02.006}
{\href{https://arxiv.org/abs/1810.12602}{{arXiv:1810.12602}}}
{[hep-ph]}
\end{barticle}
\endbibitem

%%% 2
\bibitem[\protect\citeauthoryear{Agostinelli et~al.}{2003}]{GEANT4:2002zbu}
\begin{barticle}
\bauthor{\bsnm{Agostinelli}, \binits{S.}}, \betal:
\batitle{{GEANT4 - A Simulation Toolkit}}.
\bjtitle{Nucl. Instrum. Meth. A}
\bvolume{506},
\bfpage{250}--\blpage{303}
(\byear{2003})
\doiurl{10.1016/S0168-9002(03)01368-8}
\end{barticle}
\endbibitem

%%% 3
\bibitem[\protect\citeauthoryear{Bertholet and
  Soffer}{2024}]{bertholet_soffer_2024}
\begin{botherref}
\oauthor{\bsnm{Bertholet}, \binits{E.}},
\oauthor{\bsnm{Soffer}, \binits{A.}}:
B2TrEst: charged particle reconstruction efficiency estimator for the Belle II
  experiment (v1.1.0).
Zenodo
(2024).
\doiurl{10.5281/zenodo.13831999} .
\url{https://doi.org/10.5281/zenodo.13831999}
\end{botherref}
\endbibitem

%%% 4
\bibitem[\protect\citeauthoryear{Abe et~al.}{2010}]{Belle-II:2010dht}
\begin{botherref}
\oauthor{\bsnm{Abe}, \binits{T.}}, et al.:
{Belle II Technical Design Report}
(2010)
{\href{https://arxiv.org/abs/1011.0352}{{arXiv:1011.0352}}}
{[physics.ins-det]}
\end{botherref}
\endbibitem

%%% 5
\bibitem[\protect\citeauthoryear{Altmannshofer et~al.}{2019}]{Belle-II:2018jsg}
\begin{barticle}
\bauthor{\bsnm{Altmannshofer}, \binits{W.}}, \betal:
\batitle{{The Belle II Physics Book}}.
\bjtitle{PTEP}
\bvolume{2019}(\bissue{12}),
\bfpage{123}--\blpage{01}
(\byear{2019})
\doiurl{10.1093/ptep/ptz106}
{\href{https://arxiv.org/abs/1808.10567}{{arXiv:1808.10567}}}
{[hep-ex]}.
\bcomment{[Erratum: PTEP 2020, 029201 (2020)]}
\end{barticle}
\endbibitem

%%% 6
\bibitem[\protect\citeauthoryear{Adachi et~al.}{2023}]{Belle-II:2023ueh}
\begin{barticle}
\bauthor{\bsnm{Adachi}, \binits{I.}}, \betal:
\batitle{{Search for a long-lived spin-0 mediator in b\textrightarrow{}s
  transitions at the Belle II experiment}}.
\bjtitle{Phys. Rev. D}
\bvolume{108}(\bissue{11}),
\bfpage{111104}
(\byear{2023})
\doiurl{10.1103/PhysRevD.108.L111104}
{\href{https://arxiv.org/abs/2306.02830}{{arXiv:2306.02830}}}
{[hep-ex]}
\end{barticle}
\endbibitem

%%% 7
\bibitem[\protect\citeauthoryear{Lees et~al.}{2015}]{BaBar:2015jvu}
\begin{barticle}
\bauthor{\bsnm{Lees}, \binits{J.P.}}, \betal:
\batitle{{Search for Long-Lived Particles in $e^+e^-$ Collisions}}.
\bjtitle{Phys. Rev. Lett.}
\bvolume{114}(\bissue{17}),
\bfpage{171801}
(\byear{2015})
\doiurl{10.1103/PhysRevLett.114.171801}
{\href{https://arxiv.org/abs/1502.02580}{{arXiv:1502.02580}}}
{[hep-ex]}
\end{barticle}
\endbibitem

%%% 8
\bibitem[\protect\citeauthoryear{Liventsev et~al.}{2013}]{Belle:2013ytx}
\begin{barticle}
\bauthor{\bsnm{Liventsev}, \binits{D.}}, \betal:
\batitle{{Search for heavy neutrinos at Belle}}.
\bjtitle{Phys. Rev. D}
\bvolume{87}(\bissue{7}),
\bfpage{071102}
(\byear{2013})
\doiurl{10.1103/PhysRevD.87.071102}
{\href{https://arxiv.org/abs/1301.1105}{{arXiv:1301.1105}}}
{[hep-ex]}.
\bcomment{[Erratum: Phys.Rev.D 95, 099903 (2017)]}
\end{barticle}
\endbibitem

%%% 9
\bibitem[\protect\citeauthoryear{Nayak et~al.}{2024}]{Belle:2024wyk}
\begin{botherref}
\oauthor{\bsnm{Nayak}, \binits{M.}}, et al.:
{Search for a heavy neutral lepton that mixes predominantly with the tau
  neutrino}
(2024)
{\href{https://arxiv.org/abs/2402.02580}{{arXiv:2402.02580}}}
{[hep-ex]}
\end{botherref}
\endbibitem

%%% 10
\bibitem[\protect\citeauthoryear{Dey et~al.}{2021}]{Dey:2020juy}
\begin{barticle}
\bauthor{\bsnm{Dey}, \binits{S.}},
\bauthor{\bsnm{Dib}, \binits{C.O.}},
\bauthor{\bsnm{Carlos~Helo}, \binits{J.}},
\bauthor{\bsnm{Nayak}, \binits{M.}},
\bauthor{\bsnm{Neill}, \binits{N.A.}},
\bauthor{\bsnm{Soffer}, \binits{A.}},
\bauthor{\bsnm{Wang}, \binits{Z.S.}}:
\batitle{{Long-lived light neutralinos at Belle II}}.
\bjtitle{JHEP}
\bvolume{02},
\bfpage{211}
(\byear{2021})
\doiurl{10.1007/JHEP02(2021)211}
{\href{https://arxiv.org/abs/2012.00438}{{arXiv:2012.00438}}}
{[hep-ph]}
\end{barticle}
\endbibitem

%%% 11
\bibitem[\protect\citeauthoryear{Bertholet et~al.}{2022}]{Bertholet:2021hjl}
\begin{barticle}
\bauthor{\bsnm{Bertholet}, \binits{E.}},
\bauthor{\bsnm{Chakraborty}, \binits{S.}},
\bauthor{\bsnm{Loladze}, \binits{V.}},
\bauthor{\bsnm{Okui}, \binits{T.}},
\bauthor{\bsnm{Soffer}, \binits{A.}},
\bauthor{\bsnm{Tobioka}, \binits{K.}}:
\batitle{{Heavy QCD axion at Belle II: Displaced and prompt signals}}.
\bjtitle{Phys. Rev. D}
\bvolume{105}(\bissue{7}),
\bfpage{071701}
(\byear{2022})
\doiurl{10.1103/PhysRevD.105.L071701}
{\href{https://arxiv.org/abs/2108.10331}{{arXiv:2108.10331}}}
{[hep-ph]}
\end{barticle}
\endbibitem

%%% 12
\bibitem[\protect\citeauthoryear{Dib et~al.}{2020}]{Dib:2019tuj}
\begin{barticle}
\bauthor{\bsnm{Dib}, \binits{C.O.}},
\bauthor{\bsnm{Helo}, \binits{J.C.}},
\bauthor{\bsnm{Nayak}, \binits{M.}},
\bauthor{\bsnm{Neill}, \binits{N.A.}},
\bauthor{\bsnm{Soffer}, \binits{A.}},
\bauthor{\bsnm{Zamora-Saa}, \binits{J.}}:
\batitle{{Searching for a sterile neutrino that mixes predominantly with
  $\nu_\tau$ at $B$ factories}}.
\bjtitle{Phys. Rev. D}
\bvolume{101}(\bissue{9}),
\bfpage{093003}
(\byear{2020})
\doiurl{10.1103/PhysRevD.101.093003}
{\href{https://arxiv.org/abs/1908.09719}{{arXiv:1908.09719}}}
{[hep-ph]}
\end{barticle}
\endbibitem

%%% 13
\bibitem[\protect\citeauthoryear{Ferber et~al.}{2022}]{Ferber:2022ewf}
\begin{barticle}
\bauthor{\bsnm{Ferber}, \binits{T.}},
\bauthor{\bsnm{Garcia-Cely}, \binits{C.}},
\bauthor{\bsnm{Schmidt-Hoberg}, \binits{K.}}:
\batitle{{BelleII sensitivity to long\textendash{}lived dark photons}}.
\bjtitle{Phys. Lett. B}
\bvolume{833},
\bfpage{137373}
(\byear{2022})
\doiurl{10.1016/j.physletb.2022.137373}
{\href{https://arxiv.org/abs/2202.03452}{{arXiv:2202.03452}}}
{[hep-ph]}
\end{barticle}
\endbibitem

%%% 14
\bibitem[\protect\citeauthoryear{Zhou et~al.}{2022}]{Zhou:2021ylt}
\begin{barticle}
\bauthor{\bsnm{Zhou}, \binits{G.}},
\bauthor{\bsnm{G\"unther}, \binits{J.Y.}},
\bauthor{\bsnm{Wang}, \binits{Z.S.}},
\bauthor{\bsnm{Vries}, \binits{J.}},
\bauthor{\bsnm{Dreiner}, \binits{H.K.}}:
\batitle{{Long-lived sterile neutrinos at Belle II in effective field theory}}.
\bjtitle{JHEP}
\bvolume{04},
\bfpage{057}
(\byear{2022})
\doiurl{10.1007/JHEP04(2022)057}
{\href{https://arxiv.org/abs/2111.04403}{{arXiv:2111.04403}}}
{[hep-ph]}
\end{barticle}
\endbibitem

%%% 15
\bibitem[\protect\citeauthoryear{Kang et~al.}{2021}]{Kang:2021oes}
\begin{barticle}
\bauthor{\bsnm{Kang}, \binits{D.W.}},
\bauthor{\bsnm{Ko}, \binits{P.}},
\bauthor{\bsnm{Lu}, \binits{C.-T.}}:
\batitle{{Exploring properties of long-lived particles in inelastic dark matter
  models at Belle II}}.
\bjtitle{JHEP}
\bvolume{04},
\bfpage{269}
(\byear{2021})
\doiurl{10.1007/JHEP04(2021)269}
{\href{https://arxiv.org/abs/2101.02503}{{arXiv:2101.02503}}}
{[hep-ph]}
\end{barticle}
\endbibitem

%%% 16
\bibitem[\protect\citeauthoryear{Duerr et~al.}{2021}]{Duerr:2020muu}
\begin{barticle}
\bauthor{\bsnm{Duerr}, \binits{M.}},
\bauthor{\bsnm{Ferber}, \binits{T.}},
\bauthor{\bsnm{Garcia-Cely}, \binits{C.}},
\bauthor{\bsnm{Hearty}, \binits{C.}},
\bauthor{\bsnm{Schmidt-Hoberg}, \binits{K.}}:
\batitle{{Long-lived Dark Higgs and Inelastic Dark Matter at Belle II}}.
\bjtitle{JHEP}
\bvolume{04},
\bfpage{146}
(\byear{2021})
\doiurl{10.1007/JHEP04(2021)146}
{\href{https://arxiv.org/abs/2012.08595}{{arXiv:2012.08595}}}
{[hep-ph]}
\end{barticle}
\endbibitem

%%% 17
\bibitem[\protect\citeauthoryear{Duerr et~al.}{2020}]{Duerr:2019dmv}
\begin{barticle}
\bauthor{\bsnm{Duerr}, \binits{M.}},
\bauthor{\bsnm{Ferber}, \binits{T.}},
\bauthor{\bsnm{Hearty}, \binits{C.}},
\bauthor{\bsnm{Kahlhoefer}, \binits{F.}},
\bauthor{\bsnm{Schmidt-Hoberg}, \binits{K.}},
\bauthor{\bsnm{Tunney}, \binits{P.}}:
\batitle{{Invisible and displaced dark matter signatures at Belle II}}.
\bjtitle{JHEP}
\bvolume{02},
\bfpage{039}
(\byear{2020})
\doiurl{10.1007/JHEP02(2020)039}
{\href{https://arxiv.org/abs/1911.03176}{{arXiv:1911.03176}}}
{[hep-ph]}
\end{barticle}
\endbibitem

%%% 18
\bibitem[\protect\citeauthoryear{Filimonova et~al.}{2020}]{Filimonova:2019tuy}
\begin{barticle}
\bauthor{\bsnm{Filimonova}, \binits{A.}},
\bauthor{\bsnm{Sch\"afer}, \binits{R.}},
\bauthor{\bsnm{Westhoff}, \binits{S.}}:
\batitle{{Probing dark sectors with long-lived particles at BELLE II}}.
\bjtitle{Phys. Rev. D}
\bvolume{101}(\bissue{9}),
\bfpage{095006}
(\byear{2020})
\doiurl{10.1103/PhysRevD.101.095006}
{\href{https://arxiv.org/abs/1911.03490}{{arXiv:1911.03490}}}
{[hep-ph]}
\end{barticle}
\endbibitem

%%% 19
\bibitem[\protect\citeauthoryear{Jaeckel and Phan}{2023}]{Jaeckel:2023huy}
\begin{botherref}
\oauthor{\bsnm{Jaeckel}, \binits{J.}},
\oauthor{\bsnm{Phan}, \binits{A.V.}}:
{Searching Dark Photons using displaced vertices at Belle II -- with
  backgrounds}
(2023)
{\href{https://arxiv.org/abs/2312.12522}{{arXiv:2312.12522}}}
{[hep-ph]}
\end{botherref}
\endbibitem

%%% 20
\bibitem[\protect\citeauthoryear{Bandyopadhyay
  et~al.}{2022}]{Bandyopadhyay:2022klg}
\begin{barticle}
\bauthor{\bsnm{Bandyopadhyay}, \binits{T.}},
\bauthor{\bsnm{Chakraborty}, \binits{S.}},
\bauthor{\bsnm{Trifinopoulos}, \binits{S.}}:
\batitle{{Displaced searches for light vector bosons at Belle II}}.
\bjtitle{JHEP}
\bvolume{05},
\bfpage{141}
(\byear{2022})
\doiurl{10.1007/JHEP05(2022)141}
{\href{https://arxiv.org/abs/2203.03280}{{arXiv:2203.03280}}}
{[hep-ph]}
\end{barticle}
\endbibitem

%%% 21
\bibitem[\protect\citeauthoryear{Ferber et~al.}{2023}]{Ferber:2022rsf}
\begin{barticle}
\bauthor{\bsnm{Ferber}, \binits{T.}},
\bauthor{\bsnm{Filimonova}, \binits{A.}},
\bauthor{\bsnm{Sch\"afer}, \binits{R.}},
\bauthor{\bsnm{Westhoff}, \binits{S.}}:
\batitle{{Displaced or invisible? ALPs from B decays at Belle II}}.
\bjtitle{JHEP}
\bvolume{04},
\bfpage{131}
(\byear{2023})
\doiurl{10.1007/JHEP04(2023)131}
{\href{https://arxiv.org/abs/2201.06580}{{arXiv:2201.06580}}}
{[hep-ph]}
\end{barticle}
\endbibitem

%%% 22
\bibitem[\protect\citeauthoryear{Adachi et~al.}{2024}]{Belle-II:2023ksq}
\begin{barticle}
\bauthor{\bsnm{Adachi}, \binits{I.}}, \betal:
\batitle{{Measurement of branching fractions and direct CP asymmetries for
  B\textrightarrow{}K\ensuremath{\pi} and
  B\textrightarrow{}\ensuremath{\pi}\ensuremath{\pi} decays at Belle II}}.
\bjtitle{Phys. Rev. D}
\bvolume{109}(\bissue{1}),
\bfpage{012001}
(\byear{2024})
\doiurl{10.1103/PhysRevD.109.012001}
{\href{https://arxiv.org/abs/2310.06381}{{arXiv:2310.06381}}}
{[hep-ex]}
\end{barticle}
\endbibitem

%%% 23
\bibitem[\protect\citeauthoryear{Bertacchi
  et~al.}{2021}]{BelleIITrackingGroup:2020hpx}
\begin{barticle}
\bauthor{\bsnm{Bertacchi}, \binits{V.}}, \betal:
\batitle{{Track finding at Belle II}}.
\bjtitle{Comput. Phys. Commun.}
\bvolume{259},
\bfpage{107610}
(\byear{2021})
\doiurl{10.1016/j.cpc.2020.107610}
{\href{https://arxiv.org/abs/2003.12466}{{arXiv:2003.12466}}}
{[physics.ins-det]}
\end{barticle}
\endbibitem

%%% 24
\bibitem[\protect\citeauthoryear{Reuter et~al.}{2024}]{Reuter:2024kja}
\begin{botherref}
\oauthor{\bsnm{Reuter}, \binits{L.}}, et al.:
{End-to-End Multi-Track Reconstruction using Graph Neural Networks at Belle II}
(2024)
{\href{https://arxiv.org/abs/2411.13596}{{arXiv:2411.13596}}}
{[physics.ins-det]}
\end{botherref}
\endbibitem

%%% 25
\bibitem[\protect\citeauthoryear{Sjostrand et~al.}{2006}]{Sjostrand:2006za}
\begin{barticle}
\bauthor{\bsnm{Sjostrand}, \binits{T.}},
\bauthor{\bsnm{Mrenna}, \binits{S.}},
\bauthor{\bsnm{Skands}, \binits{P.Z.}}:
\batitle{{PYTHIA 6.4 Physics and Manual}}.
\bjtitle{JHEP}
\bvolume{05},
\bfpage{026}
(\byear{2006})
\doiurl{10.1088/1126-6708/2006/05/026}
{\href{https://arxiv.org/abs/hep-ph/0603175}{{arXiv:hep-ph/0603175}}}
\end{barticle}
\endbibitem

%%% 26
\bibitem[\protect\citeauthoryear{Lange}{2001}]{Lange:2001uf}
\begin{barticle}
\bauthor{\bsnm{Lange}, \binits{D.J.}}:
\batitle{{The EvtGen particle decay simulation package}}.
\bjtitle{Nucl. Instrum. Meth. A}
\bvolume{462},
\bfpage{152}--\blpage{155}
(\byear{2001})
\doiurl{10.1016/S0168-9002(01)00089-4}
\end{barticle}
\endbibitem

%%% 27
\bibitem[\protect\citeauthoryear{{Belle II Collaboration}}{2020}]{BelleIIeff}
\begin{botherref}
\oauthor{\bsnm{{Belle II Collaboration}}}:
Measurement of the tracking efficiency and fake rate with $e^+e^- \to
  \tau^+\tau^-$ events.
Report No. BELLE2-NOTE-PL-2020-014
(2020)
\end{botherref}
\endbibitem

%%% 28
\bibitem[\protect\citeauthoryear{Allmendinger
  et~al.}{2013}]{Allmendinger:2012ch}
\begin{barticle}
\bauthor{\bsnm{Allmendinger}, \binits{T.}}, \betal:
\batitle{{Track Finding Efficiency in BaBar}}.
\bjtitle{Nucl. Instrum. Meth. A}
\bvolume{704},
\bfpage{44}--\blpage{59}
(\byear{2013})
\doiurl{10.1016/j.nima.2012.11.184}
{\href{https://arxiv.org/abs/1207.2849}{{arXiv:1207.2849}}}
{[hep-ex]}
\end{barticle}
\endbibitem

%%% 29
\bibitem[\protect\citeauthoryear{Alcaraz~Maestre}{2021}]{AlcarazMaestre:2021oeh}
\begin{botherref}
\oauthor{\bsnm{Alcaraz~Maestre}, \binits{J.}}:
{Tracking charged particles in the zero curvature limit}
(2021)
{\href{https://arxiv.org/abs/2109.00845}{{arXiv:2109.00845}}}
{[physics.ins-det]}
\end{botherref}
\endbibitem

%%% 30
\bibitem[\protect\citeauthoryear{Krohn
  et~al.}{2020}]{Belle-IIanalysissoftwareGroup:2019dlq}
\begin{barticle}
\bauthor{\bsnm{Krohn}, \binits{J.-F.}}, \betal:
\batitle{{Global decay chain vertex fitting at Belle II}}.
\bjtitle{Nucl. Instrum. Meth. A}
\bvolume{976},
\bfpage{164269}
(\byear{2020})
\doiurl{10.1016/j.nima.2020.164269}
{\href{https://arxiv.org/abs/1901.11198}{{arXiv:1901.11198}}}
{[hep-ex]}
\end{barticle}
\endbibitem

\end{thebibliography}

\end{document}